\documentclass[twocolumn,showpacs,pre,aps,superscriptaddress]{revtex4}
\usepackage[latin1]{inputenc}
\usepackage{bm}
\usepackage{multirow}
\usepackage{amssymb}
\usepackage{amsbsy}
\usepackage{amsmath}
\usepackage{stmaryrd}
\usepackage{graphicx}
\usepackage{epsfig}

\begin{document}

\title{Projection operator approach to master equations for coarse grained occupation numbers in non-ideal quantum gases }

\author{Christian Bartsch}

\email{cbartsch@uos.de}

\affiliation{Fachbereich Physik, Universit\"at Osnabr\"uck,
             Barbarastrasse 7, D-49069 Osnabr\"uck, Germany}

\author{Robin Steinigeweg}

\email{r.steinigeweg@tu-bs.de}

\affiliation{Institut f\"ur Theoretische Physik, Technische Universit\"at Braunschweig,
            Mendelsohnstrasse 3, D-38106 Braunschweig, Germany}
   
\author{Jochen Gemmer}

\email{jgemmer@uos.de}

\affiliation{Fachbereich Physik, Universit\"at Osnabr\"uck,
             Barbarastrasse 7, D-49069 Osnabr\"uck, Germany}

\date{\today}

\begin{abstract}
We aim at deriving an equation of motion for specific sums of momentum mode occupation numbers from models for electrons in periodic lattices experiencing elastic scattering, electron-phonon scattering or electron-electron scattering. These sums correspond to ``grains'' in momentum space. This equation of motion is supposed to involve only a moderate number of dynamical variables and/or exhibit a sufficiently simple structure such that neither its construction nor its analyzation/solution requires substantial numerical effort. To this we end compute, by means of a projection operator technique, a linear(ized) collision term which determines the dynamics of the above grain-sums. This collision term results as non-singular, finite dimensional rate matrix and may thus be inverted regardless of any symmetry of the underlying model. This facilitates calculations of, e.g., transport coefficients, as we demonstrate for a 3-dim. Anderson model featuring weak disorder.
\end{abstract}

\pacs{
05.30.-d, 
05.70.Ln,  
72.10.-d,  
05.60.Gg  
}

\maketitle

%
%

\section{Introduction} \label{sec-introduction}
The mapping of occupation number dynamics of non-ideal quantum gases (as resulting from the Schroedinger equation) onto Boltzmann equations is a classical topic in the literature \cite{mertig1987,ziman1960,jaeckle1978,banyai2006}. As well-known the corresponding approaches routinely rely on approximations and/or conceptual assumptions of some sort (for an overview see App.~\ref{sec-app}). Furthermore, they often yield rather formal expressions for the collision terms involving, e.g, singular scattering rates and a diverging number of occupation numbers for a description based on a discrete reciprocal lattice. If the singularities are regularized and the description is transferred to a continuous momentum space, the collision term results as a (possibly non-linear) map of a scalar function onto another scalar function. For many purposes such as calculation of transport coefficients, electronic lifetimes  etc. eigenfunctions and eigenvalues of this map are needed. Those are, however, only easy to obtain for full spherical symmetry of the model, i.e., isotropic scattering and fully isotropic dispersion relations. If strict spherical symmetry is absent, many approaches resort to an approximate ``transport lifetime'' which essentially amounts to an educated guess of the relevant eigenvalues \cite{jaeckle1978}.  Presumably due to this subtleties detailed calculations of, e.g., transport coefficients for models featuring (massively) anisotropic dispersion relations are rare in the literature.

In this paper we thus suggest an alternative construction of collision terms from pertinent underlying quantum models. These collision terms directly result as finite dimensional, non-singular rate matrices. Their construction as well as their numerical inversion may be performed on standard desktop computers. No singularities arise in the derivation of these collision terms. The approach is based on a (optimizeable) ``coarse graining'' in momentum space. The above rates may be interpreted as electronic transition rates between these grains. A carefully chosen graining may yield a particularly favorable form of the collision term, regardless of any spherical symmetry, etc., as will be demonstrated below by the computation of diffusion coefficients for an Anderson model, cf. Sec.~\ref{sec-dca}. Our approach is an implementation of  the time-convolutionless projection operator method (TCL). The latter is routinely used to find an autonomous equation of motion for a set of variables of interest. Those are here essentially the accumulated occupation numbers of the grains.  Within the frame of the TCL approach no possibly ill-controlled, additional assumptions (molecular chaos, full factorizability, etc.) have to be employed. However, in the paper at hand we concentrate on a leading order truncation of the TCL-expansion (cf. (\ref{expva})). (Whether or not this will yield sufficiently accurate results is a somewhat subtle question but can in principle be checked by explicit evaluation of higher order terms \cite{bartsch2008}.) For a discussion of the relation of the present results to the afore mentioned traditional approaches, see paragraph below (\ref{rateimp}).

Our paper is organized as follows: First (Sec.~\ref{sec-tclp}) we give a very brief introduction to the TCL projection operator formalism. In Sec.~\ref{sec-smpp} we introduce some archetypical condensed matter models which are routinely used to investigate non-ideal electronic quantum gases as arising from elastic scattering, electron-phonon interaction and electron-electron interaction. We furthermore specify our projection operator which is especially constructed to investigate the dynamics of the above mentioned, occupation number related variables. Then (Sec.~\ref{sec-application}) we perform leading order calculations on these models. Those result in general expressions for the transition rates which may be further evaluated on the basis of concrete dispersion relations and scattering potentials. This is done in (Sec.~\ref{sec-dca}) for a weak disorder $3$-d Anderson model. It is thereby demonstrated how the graining can be chosen to cast the collision term in a form for which a relaxation-time approximation actually becomes exact. This facilitates the simple calculation of an energy-dependent diffusion coefficient. The latter significantly differs from a result obtained from an approach based on full spherical symmetry.

%
%

\section{Time-convolutionless Projective approach to relevant dynamics} \label{sec-tclp}

In this paragraph we give a short introduction to the TCL projection operator method \cite{chaturvedi1979,breuer2006,breuer2007}. In general, the latter is a perturbative projection
operator technique which describes (the) reduced dynamics of a quantum system with a Hamiltonian of the type $H = H_{0} + \lambda V$, where $\lambda V$ has to be small in some sense. 
It produces autonomous equations of motion for the variables of interest (``relevant information'').
In order to apply this method one first has to construct a suitable
projection (super) operator $\mathcal{P}$ which projects any density matrix $\rho(t)$ onto its relevant part. ``Relevant part'' here implies that   $\mathcal{P}\rho(t)$, in spite of being significantly less complex than $\rho(t)$, should still contain all variables of interest. Furthermore, $\mathcal{P}$ has to satisfy the characteristic trait of a projection operator
$\mathcal{P}^{2} \,\rho(t) = \mathcal{P} \,\rho(t)$.
For initial states with $\mathcal{P}\rho (0)=\rho (0)$ the TCL scheme yields a closed time-local differential equation for the dynamics of 
$\mathcal{P}\rho$,

\begin{eqnarray}
\frac{\partial}{\partial t}\mathcal{P}\rho(t)=\mathcal{K}(t)\mathcal{P}\rho(t), \hspace{0.5cm} \mathcal{K}(t)
=\sum\limits_{i=1}^{\infty}\lambda^{i}\mathcal{K}_{i}(t),
\label{expva}
\end{eqnarray}
where the perturbative expansion used in the last equations is in
principle (formally) exact. In this paper we exclusively focus on initial states which satisfy the above relation. For a discussion of the legitimacy of this approach see \cite{grabert1977,breuer2007,romero2004,bartsch2009}.

As already mentioned in the introduction, we furthermore focus on a description to leading order of (\ref{expva}).
which is typically and certainly in our case the second order, i.e., we have to
determine $\mathcal{K}_{2}(t)$. A widely accepted
indicator  for the validity of the truncation is a clear timescale
separation between the resulting relaxation dynamics and the decay of
the correlation function, the latter being introduced below.

In the
literature \cite{breuer2007} one finds 

\begin{equation}
\mathcal{K}_{2}(t) = \int_{0}^{t} \text{d}t_1 \, \mathcal{P} \,
\mathcal{L}(t) \, \mathcal{L}(t_1) \, \mathcal{P} \; .
\end{equation}
where $\mathcal{L}(t)\rho=-\frac{\imath}{\hbar}[V(t),\rho]$ corresponds
to the Liouville- (super-) operator.
Here and in the following all equations are denoted in the
interaction picture.
For a concrete application we have to specify the underlying
quantum model and further a suitable projection operator.

\section{Structure of the models and a pertinent projection operator}
\label{sec-smpp}
The systems we discuss here may in general be all sorts of quantum gases, but for clarity and brevity we focus  on  the ``spinless fermions''- type here. (This refers to the particles for which the collision term is to be constructed.)
\begin{eqnarray}
H=\underbrace{\sum\limits_{{\bf k}}\varepsilon_{{\bf k}}a_{{\bf k}}^{\dagger}a_{{\bf 
k}}}_{H_{0}}+V
\label{Hamilton1}
\end{eqnarray}
$a_{{\bf k}}^{\dagger},a_{{\bf k}}$ are electronic creation/annihilation operators
in some momentum eigenmodes and $\varepsilon_{{\bf k}}$ denotes the corresponding
dispersion relation. 
The latter depends on the underlying model which is assumed to describe the non-interacting electrons, e.g., a free electron gas,
a tight binding model, etc.. An adequate dispersion relation may also be determined from an advanced solid state method such as density functional theory, etc..

Here, $V$ refers to different pertinent interaction types, which we specify and investigate below in paragraph \ref{sec-application}. In detail, these are elastic scattering, electron-phonon-interaction and electron-electron-interaction. All of them are treated as small perturbations
(in the sense of the TCL method).

Now, in order to introduce a pertinent projection operator, we firstly need to define some basic operators.
For the non-interacting 
many-particle system we may directly write down the
wavenumber (momentum) dependent ``single particle equilibrium density operator'' on the mode ${\bf j}$ as:
\begin{eqnarray}
\rho_{{\bf j}}^{\text{eq}}:=f_{{\bf j}}(\mu,T)a_{\bf j}^{\dagger}a_{\bf j}+(1-f_{{\bf j}}(\mu,T))a_{\bf j}a_{\bf j}^{\dagger},
\end{eqnarray}
with $f_{{\bf j}}(\mu,T)=(\exp ((\varepsilon({\bf j})-\mu)/k_{B}T)+1)^{-1}$ being the Fermi distribution.
Since we are interested in low temperature 
regimes we may substitute (approximate) the chemical potential $\mu$  by the Fermi energy $\varepsilon_{F}$. Further we abbreviate $f_{{\bf 
j}}(\varepsilon_{F},T)$ as $f_{{\bf j}}$.

The equilibrium density operator, again for the non-interacting case (, i.e., $H_0$),
of the total system, $\rho^{\text{eq}}$, may be written as the tensor product of the single particle 
density operators, i.e.,
\begin{eqnarray}
\rho^{\text{eq}}:=\bigotimes\limits_{{\bf i}}\rho_{{\bf
    i}}^{\text{eq}}.
\label{density}
\end{eqnarray}

We should mention here for later reference that, while $\rho^{\text{eq}}$ is strictly speaking
just the equilibrium state of the non-interacting system, it is
routinely considered to describe the equilibrium single particle properties of the
weakly interacting system more or less correctly. Thus, if single
particle observables  
relax towards equilibrium due to the
interactions (scattering), we expect them to relax towards values
corresponding to $\rho^{\text{eq}}$.\\


Furthermore, the deviation of the mode occupation number $n_{\bf j}=a_{\bf j}^{\dagger}a_{\bf j}$ from its thermal equilibrium $f_{{\bf j}}$ may be described by an operator $\Delta_{{\bf j}}$ which we define as
\begin{eqnarray}
\Delta_{{\bf j}}:=(1-f_{{\bf j}})a_{\bf j}^{\dagger}a_{\bf j}-f_{{\bf j}}a_{\bf j}a_{\bf j}^{\dagger}=
a_{\bf j}^{\dagger}a_{\bf j}-f_{{\bf j}}. 
\end{eqnarray}

In order to concretize the projector further we must now introduce the afore mentioned coarse graining in momentum space. Eventually this means that we have to define domains in momentum space; we label those regions by Greek indices ($\kappa, \eta$). Different occupation numbers corresponding to the same region will no longer be investigated separately from each other, the remaining variables will simply be the sums over occupation numbers belonging to the various grains. The concrete choice of the grains substantially influences the result of the following considerations, as will become clear in the remainder of this paper. However, at this point we simply mathematically define operators $\Delta^{\kappa}$ describing the summed deviations (from equilibrium) of all occupation numbers belonging to a common grain

\begin{equation}
\Delta^{\kappa}=\sum_{{\bf j}\in\kappa} \Delta_{{\bf j}} \ . 
\end{equation}
Furthermore, we define operators
\begin{equation}
D_{{\bf j}}:=\frac{\rho^{eq}\Delta_{{\bf j}}}{\text{Tr}\{\rho^{eq}\Delta^{2}_{{\bf 
j}}\}}=
\bigotimes\limits_{{\bf i\neq {\bf j}}}\rho_{{\bf
    i}}^{\text{eq}} (a_{\bf j}^{\dagger}a_{\bf j}-a_{\bf j}a_{\bf j}^{\dagger})
\label{Ddef}
\end{equation}
$D_{\bf j}$ corresponds to a diagonal product operator that equals
the equilibrium density operator on all modes except ${\bf j}$, 
and essentially the occupation number operator (up to some substracted unity operator) on the mode ${\bf j}$.
The above definition of $D_{\bf j}$ ensures that the important relation
\begin{equation}
\text{Tr}\{ D_{{\bf j}}\Delta_{{\bf l}}\}=\delta_{\bf j,l}
\label{deltas}
\end{equation}
is fulfilled, which turns out to be crucial for the idempotency property of the below defined projection operator.

Moreover, we abbreviate the time dependent expectation value of $\Delta^{\kappa}$, which we are mainly interested in by $d^{\kappa}(t):=$Tr$\{\Delta^{\kappa}\rho(t)\}$,
where $\rho(t)$ is the density operator which describes the actual state of the system.

With these preliminary definitions we construct a suitable projector as follows:
\begin{eqnarray}
\mathcal{P}\rho(t)=\text{Tr}\{ \rho(t)\}\rho^{\text{eq}}+\sum_{\kappa}\frac{1}{N_{\kappa}}(\sum_{{\bf j}\in\kappa}D_{{\bf j}}) d^{\kappa}(t)
\label{proj}
\end{eqnarray}
Here $N_{\kappa}$ denotes the number of momentum states in grain $\kappa$. In the following we exclusively consider states satisfying $\text{Tr}\{ \rho(t)\} = 1$. (Of course this trace is invariant under unitary time evolution.)

In general, the relevant part of the density matrix $\mathcal{P}\rho(t)$ is a non-factorizing state. Due to relation (\ref{Ddef}) the addition of a term containing $D_{{\bf j}}$ to the equilibrium state $\rho^{\text{eq}}$ in the construction (\ref{proj}) may be interpreted as adding a linear deviation to the equilibrium occupation number of mode ${\bf j}$, while leaving the other occupation numbers unchanged. 

Note that a similar projector has recently been used in the context of investigations on electronic lifetimes in aluminum \cite{kadiroglu2009}. In that case one is interested in describing the decay of a single electronic excitation and therefore the corresponding projector is constructed to project only onto one single occupation number as dynamical variable, whereas the projector (\ref{proj}) used in this paper keeps all grain occupation numbers as time dependent variables thus giving a more detailed picture of the dynamics. In so far the considerations in this paper extend the previous work from \cite{kadiroglu2009}. Additionally, the $ D_{{\bf j}}$ are here defined in a slightly different way which leads to a simplification of the upcoming calculations.

By exploiting (\ref{deltas}) one may straightforwardly prove that this projector indeed features the crucial idempotency property of a projection operator, i.e., $\mathcal{P}^2=\mathcal{P}$. It furthermore obviously captures the dynamical variables of interest, namely the $d^{\kappa}$.

Before we eventually concretely apply (\ref{expva}) we
make the following approximation for an expression that appears in the
computation of (\ref{expva}):
\begin{eqnarray}
\frac{\imath}{\hbar}[V(t),\rho^{\text{eq}}]\approx 0
\label{approx1}
\end{eqnarray}
The neglected commutator term essentially describes the dynamics of the
equilibrium state of the non-interacting system. Eventually we are
interested in a single particle observable. As already mentioned
above, the equilibrium state of the non-interacting system is believed
to
reasonably describe single particle observables in equilibrium even
for weakly interacting systems. Since an equilibrium state is
constant, the above commutator should not significantly contribute to
the relevant dynamics, thus we drop it. Keeping the term and
performing all following steps  eventually yields an expression which
can be explicitly shown to be indeed negligible in the weak coupling
limit. For clarity and briefness we omit this calculation here.\\

Now, we explicitly evaluate (\ref{expva}) to leading, i.e., second order using the above projector. In order to extract from this equation of motion for operators an equation of motion for the scalar observables of interest (which are the $d^\eta(t)$) we multiply by $\Delta^\eta$ and take a trace: 
\begin{equation}
\frac{\partial}{\partial t}\text{Tr}\{ \Delta^{\eta}\mathcal{P}\rho(t)\}=\text{Tr}\{ \Delta^{\eta} \mathcal{K}_{2}(t)\mathcal{P}\rho(t)\} 
\end{equation}
Exploiting (\ref{deltas}) and
$[V(t),\Delta_{\bf k}]=[V(t),n_{\bf k}]$, furthermore employing (\ref{approx1}) and some invariance properties of the trace, we find after a lengthy but straightforward calculation:

\begin{equation}
\dot{d}^{\eta}(t)\! =\!\sum_{\kappa}\!\int_{0}^{t}\!\!\! \text{d}t_1 \underbrace{\frac{1}{\hbar^{2}N_{\kappa}} \sum_{\substack{{\bf i}\in\kappa,\\ {\bf k}\in\eta}}\! -\text{Tr}\{D_{\bf i}[V(t_1),[V(t),n_{\bf k}]]\}}_{C_{\eta\kappa}(t,t_1)} d^{\kappa}(t)
\label{boltz}
\end{equation}
This is our first main result. Obviously (\ref{boltz}) may be interpreted as a rate equation for the dynamics of the coarse grained occupation numbers. It thus corresponds to a linear(ized) collision term. The rates, which are typically finite and directly computable, are given by $R_{\eta\kappa}(t):=\int_{0}^{t} \text{d}t_1 C_{\eta\kappa}(t,t_1)$. They may in general be time-dependent which does not fit into the standard picture. But, as they are given by integrals over correlation functions, they can be expected to converge to constant values after some correlation times under some rather mild conditions on the model and the graining (, cf. also paragraph below (\ref{rateimp})). For a concrete example for the evaluation of the rates see Sec.\ref{sec-dca}.

For most routinely considered interactions (and all interactions analyzed in this paper) the particle number within the specifically addressed system (e.g., electron system) ($N=\sum_{{\bf k}}n_{\bf k}$) is conserved. Thus 
\begin{equation}
\sum_{{\bf k}\in\eta}n_{\bf k}=N-\sum_{\kappa\neq\eta}\sum_{{\bf l}\in\kappa}n_{\bf l}
\label{cons}
\end{equation}
holds. Exploiting this feature, we may determine the diagonal rate terms of the rate equation (\ref{boltz}) from the non-diagonal terms by inserting (\ref{cons}) into (\ref{boltz}) 
\begin{equation}
R_{\eta\eta}=-\sum_{\kappa\neq\eta}R_{\kappa\eta}\ . 
\label{diag}
\end{equation}
Thus, the rate equation (\ref{boltz}) may be classified as a master equation which is consistent with its interpretation as a collision term in a Boltzmann equation.

For the following calculations it turns out to be convenient to reformulate  the trace term from (\ref{boltz}):
\begin{eqnarray}
&&-\text{Tr}\{D_{\bf i}[V(t_1),[V(t),n_{\bf k}]]\}= \nonumber\\
&&-(\text{Tr}\{ V(t)V(t_1)n_{\bf k}D_{\bf i}\}+c.c.)\nonumber\\
&&+(\text{Tr}\{ V(t)n_{\bf k}V(t_1)D_{\bf i}\}+c.c.) \ ,
\label{nocom}
\end{eqnarray}

where we have used that $D_{\bf i}$ commutes with $n_{\bf k}$.  
Note that the Hermitian conjugate is of identical form, respectively, 
except exchanged time arguments.

The physical implications of (\ref{boltz}) may become more transparent if it is applied to concrete models and some pertinent interactions are inserted. This is done for some examples in the next Section.

\section{Application to models featuring different interactions}\label{sec-application}

\subsection{Elastic Scattering}
\label{sec-elimp}
Firstly, we apply the method introduced above to a quantum gas model according to which, in addition to a periodic crystal lattice potential, the electron is subject to a weak, non-periodic potential. The latter of course induces the scattering. (One may think here for example of a few impurity atoms in the lattice of a metal or of an Anderson model with very weak on-site disorder, way down below the critical disorder, cf. \cite{grussbach1995}. The latter will be analyzed in detail in Sec.~\ref{sec-dca}.) A corresponding Hamiltonian is given by 
\begin{eqnarray}
H=\underbrace{\sum\limits_{{\bf n}}\varepsilon_{{\bf n}}a_{{\bf n}}^{\dagger}a_{{\bf 
n}}}_{H_{0, \text{el}}}+\underbrace{\sum_{{\bf k,q}}\frac{W({\bf q})}{\sqrt{\Omega}}a_{{\bf k+q}}^{\dagger}a_{{\bf k}}}_{V}\ .
\label{Hamiltoneldef}
\end{eqnarray}
Here $W({\bf q})$ simply denotes the corresponding spatial Fourier component of the impurity potential and $\Omega$ corresponds to the total number of discrete (quasi-)momenta, i.e., $\Omega$ scales with the volume of the crystal. The above ``interaction term $V$''  has to be inserted into Eq.(\ref{boltz}) to determine the scattering rates $R_{\eta\kappa}(t)$.
We evaluate the traces in (\ref{nocom}) by identifying all contributing diagonal terms, i.e., terms that contain equally many creation and annihilation operators acting on a single ${\bf k}$-mode. In doing so it proves to be advantageous
that both $D_{\bf i}$ and $n_{\bf k}$ factorize onto the single ${\bf k}$-modes and, furthermore, are diagonal on each ${\bf k}$-mode. 
After some straightforward calculation 
we finally obtain for the rates
\begin{equation}
R_{\eta\kappa}(t)=\int_{0}^{t}\!\!\!\text{d}\tau \frac{2}{\hbar^{2}N_{\kappa}}\sum_{\substack{{\bf i}\in\kappa,\\{\bf k}\in\eta}}\frac{\vert W({\bf k}-{\bf i})\vert ^{2}}{\Omega}
\cos [\frac{1}{\hbar}(\varepsilon_{{\bf i}}-\varepsilon_{{\bf k}})\tau]\ , 
\label{rateimp}
\end{equation}
with $\tau:=t-t_1$.
(Here we have used $W({\bf q})=W^{*}(-{\bf q})$, since the interaction potential should be real in configuration space).

As multiply announced, the transition rates between the grains result as non-singular, real numbers which are defined at any time
. Their concrete form obviously depends on the model as well as on the graining. Rates may become time-independent on a timescale $\propto \hbar/\Delta E$, $\Delta E$ being the energy range spanned by a grain. This together with the requirement of the dispersion relation being approximately linear within the grains sets the timescale beyond which a description based on constant rates may apply. This timescale may obviously be substantially shorter than the ``infinite time limit'' which is frequently addressed within the framework of other approaches (see App.~\ref{sec-app}). If this timescale is on the order of or shorter than a (hypothetical) relaxation time (which one would obtain from this consideration) the ``weak coupling limit'' is violated, hence no description whatsoever in terms of constant rates will apply. Furthermore, only rates between grains located within the same energy range will be non-vanishing beyond the above timescale. Rates describing transitions into a certain grain tend to increase with the corresponding cell containing more and more momentum modes. All of the above appears comparable to transition rates as obtained from Fermi's Golden Rule. Indeed, in the afore mentioned ``long time limit'' the rates from (\ref{rateimp}) assume the same values that one would get from, e.g., a ``Fermi's Golden Rule approach''(see Sec.~\ref{sec-introduction}, App.~\ref{sec-app}) after regularizing the singularities and expanding the continuous occupation number function in terms of orthonormal ``box functions'' which correspond to the grains. So in a sense our projective approach ``integrates'' the above steps of a traditional approach and puts the result on an alternative theoretical footing. This ``shortcut'' may simplify detailed calculations based on the collision term significantly, as demonstrated in Sec.~\ref{sec-dca}.

Note, that the resulting rates for this model are completely independent of any equilibrium occupation numbers, i.e., the linear dynamics do not depend on the position in momentum space at which particles have been added or taken away. Or, to rephrase, deviations from equilibrium relax all in the same way regardless of whether they occur above, below, or at the Fermi-level.

\subsection{Electron-Phonon-Interaction}

One important mechanism that is commonly believed to mainly control electronic transport in metals (at higher temperatures) is electron-phonon-scattering. To investigate this case, we routinely assume a Hamiltonian of the following form:

\begin{equation}
H\!=\!\underbrace{\sum\limits_{{\bf n}}\varepsilon_{{\bf n}}a_{{\bf n}}^{\dagger}a_{{\bf 
n}}}_{H_{0, \text{el}}}\! +\!\underbrace{\sum_{{\bf i}}\omega_{{\bf i}}b_{{\bf i}}^{\dagger}b_{{\bf 
i}}}_{H_{0, \text{ph}}}\! +\!\underbrace{(\sum_{{\bf k,q}}\frac{W({\bf q})}{\sqrt{\Omega}}a_{{\bf k+q}}^{\dagger}a_{{\bf k}}b_{{\bf q}}\! +\! h.c.)}_{V}, 
\label{Hamiltonelph}
\end{equation}

where $b_{{\bf i}}^{\dagger}, b_{{\bf i}}$ are (bosonic) creation/annihilation operators corresponding to some phonon eigenmodes (labelled by {{\bf i}}) and $\omega_{{\bf i}}$ denotes
the corresponding phononic dispersion relation.
This Hamiltonian is of the quantum gas type we have introduced above (\ref{Hamilton1}), but with an additional phononic Hamiltonian $H_{0, \text{ph}}$. The interaction $V$ may be viewed as representing processes in which an electron is scattered under the annihilation/creation of a phonon such that the complete momentum
is conserved.

We choose here to project onto deviations from equilibrium in the electronic system only, i.e., phononic occupation numbers are not treated as dynamical variables but their equilibrium values rather enter as parameters. To this end we keep the projection operator essentially as given in (\ref{proj}) but multiply $\rho^{\text{eq}}$ and $D_{\bf j}$ by $\rho^{\text{eq, Ph}}$, the latter being the equilibrium state of the phononic system (which in itself is an ideal gas). A corresponding calculation (which is essentially very similar to the previous one for electron-defect-scattering) yields
\begin{eqnarray}
&&\hspace{-0.5cm}-\text{Tr}\{D_{\bf i}[V(t_1),[V(t),n_{\bf k}]]\}= \nonumber\\
&&\hspace{-0.5cm}\frac{\vert W({\bf i}-{\bf k})\vert ^{2}}{\Omega}(f_{{\bf k}}g_{{\bf i}-{\bf k}}+(1-f_{{\bf k}})(1+g_{{\bf i}-{\bf k}}))\nonumber \\
&&\hspace{-0.5cm}\cdot 2\cos [\frac{1}{\hbar}(\varepsilon_{{\bf i}}-\varepsilon_{{\bf k}}-\omega_{{\bf i}-{\bf k}})(t-t_1)] \nonumber \\
&&\hspace{-0.5cm}+\frac{\vert W({\bf k}-{\bf i})\vert ^{2}}{\Omega}(f_{{\bf k}}(1+g_{{\bf k}-{\bf i}})+(1-f_{{\bf k}})g_{{\bf k}-{\bf i}}) \nonumber \\
&&\hspace{-0.5cm}\cdot 2\cos [\frac{1}{\hbar}(\varepsilon_{{\bf i}}-\varepsilon_{{\bf k}}+\omega_{{\bf k}-{\bf i}})(t-t_1)] \ ,
\label{eq-tr-ph}
\end{eqnarray}
where we have used $\text{Tr}\{ \rho^{\text{eq, Ph}}b_{{\bf q}}^{\dagger}b_{{\bf q}}\}=g_{{\bf q}}$.
To the corresponding rates a statement which is essentially analog to the one below (\ref{rateimp}) applies. Again, in the long time limit this rate matrix assumes a form comparable to the one which one would obtain from a traditional approach using the above mentioned ``grain-corresponding basis functions'' to represent the collision operator. However, in this case this only holds if the traditional approach is based on a linearized collision operator. So again the present approach integrates the steps of the traditional approach, including linearization. 

In many pertinent systems, the phonon energies are assumed to be small compared to the electron energies. Thus, neglecting the phonon energies ($\omega_{{\bf k}-{\bf i}}\approx 0$) 
, we may further simplify (\ref{eq-tr-ph})
and obtain for the transition rates of Eq.(\ref{boltz}) ($\kappa\neq\eta$)
\begin{eqnarray}
&&R_{\eta\kappa}(t)= \nonumber\\
&&\hspace{-0.5cm}\int_{0}^{t}\!\!\!\text{d}\tau \frac{2}{\hbar^{2}N_{\kappa}}\sum_{\substack{{\bf i}\in\kappa,\\{\bf k}\in\eta}}\frac{\vert W({\bf k}-{\bf i})\vert ^{2}}{\Omega}(1\! +\! g_{{\bf k}-{\bf i}}\! +\! g_{{\bf i}-{\bf k}})
\cos [\frac{1}{\hbar}(\varepsilon_{{\bf i}}\! -\!\varepsilon_{{\bf k}})\tau]. \nonumber\\
\end{eqnarray}
Thus, according to this approximation, the rate matrix is essentially of the same form as in the case of impurity scattering, except for the equilibrium phonon occupation numbers. Such a formulation provides (in contrast to other approaches) a directly and easily accessible starting point for massively non-isotropic systems for which any approximation based on full spherical symmetry must necessarily fail. The calculation of electronic diffusion coefficients in atomic wires as performed in \cite{bartsch2010} may serve as an example.
(For another time-convolutionless master equation approach to the dynamics of coupled electron-phonon systems see \cite{pouthier2009}.)

\subsection{Electron-Electron-Interaction}

In this paragraph we apply the above method to a system featuring electron-electron-interaction. A corresponding Hamiltonian may read
\begin{eqnarray}
H=\underbrace{\sum\limits_{{\bf n}}\varepsilon_{{\bf n}}a_{{\bf n}}^{\dagger}a_{{\bf 
n}}}_{H_{0, \text{el}}}+\underbrace{\frac{1}{2}\sum_{{\bf k,l,q}}\frac{W({\bf q})}{\Omega}a_{{\bf k+q}}^{\dagger}a_{{\bf l-q}}^{\dagger}a_{{\bf k}}a_{{\bf l}}}_{V}\ .
\label{Hamiltonelel}
\end{eqnarray}
The evaluation of the trace term (\ref{nocom}) is basically similar to the previous examples. Nevertheless, identifying all contributing terms proves in this case to be somewhat subtle. However, simply following the scheme, we arrive after some lengthy calculation at
\begin{eqnarray}
&&\hspace{-0.5cm}R_{\eta\kappa}(t)=\int_{0}^{t}\text{d}\tau \frac{2}{\hbar^{2}N_{\kappa}}\sum_{\substack{{\bf i}\in\kappa,\\{\bf k}\in\eta}}\sum_{\bf l} \nonumber\\
&&\hspace{-0.5cm}(\text{Re}(W({\bf i}-{\bf k}))-\text{Re}(W({\bf k}-{\bf l})))^{2}\cdot\frac{1}{\Omega ^{2}}\nonumber\\
&&\hspace{-0.5cm}(f_{{\bf l}}(1-f_{{\bf i-k+l}})(1-f_{{\bf k}}) 
+f_{{\bf k}}f_{{\bf i-k+l}}(1-f_{{\bf l}})
) \nonumber\\
&&\hspace{-0.5cm}\cos [\frac{1}{\hbar}(\varepsilon_{{\bf i}}+\varepsilon_{{\bf l}}-\varepsilon_{{\bf k}}-\varepsilon_{{\bf i-k+l}})\tau] \nonumber\\
&&\hspace{-0.5cm}-(\text{Re}(W({\bf l}-{\bf i}))-\text{Re}(W({\bf k}-{\bf l})))^{2}\cdot\frac{1}{\Omega ^{2}}\nonumber\\
&&\hspace{-0.5cm}(f_{{\bf k}}(1-f_{{\bf i+k-l}})(1-f_{{\bf l}})
+f_{{\bf l}}f_{{\bf i+k-l}}(1-f_{{\bf j}})
)\nonumber\\
&&\hspace{-0.5cm}\cos [\frac{1}{\hbar}(\varepsilon_{{\bf k}}+\varepsilon_{{\bf i}}-\varepsilon_{{\bf i+k-l}}-\varepsilon_{{\bf l}})\tau]  \ .
\label{el-el}
\end{eqnarray}

To those rates a statement analog to the ones below (\ref{rateimp}) and (\ref{eq-tr-ph}) applies. The present result is comparable to (and in a sense integrates) all steps of a traditional approach including linearization. The explicit possibility to choose a coarse graining in momentum space may also help here to cast the collision operator into an especially suitable form.
We intend to use (\ref{el-el}) in order to determine transport coefficients of one-dimensional models of interacting fermions as considered in \cite{sirker2009} in a forthcoming paper. While in more realistic models the relevance of electron-electron scattering for transport has to be questioned, its primary relevance for lifetimes of excited ``hot electrons'' in metals is undisputed. An approximation for the decay rate of such a hot electron results if the transition rates from one grain to all other grains in (\ref{eq-tr-ph}) are summed over. This turns out to be equivalent to employing a projector which keeps only the occupation number deviation within one grain as a relevant observable. Such an investigation has been performed (on the basis of a projector that structurally slightly differs from the one used in the present paper) in \cite{kadiroglu2009}. There reasonable lifetimes for hot electrons in aluminum have been found.

\section{Application: Diffusion coefficient of the 3-d Anderson model featuring weak disorder}
\label{sec-dca}
In this section we aim at finding the diffusion coefficient for a 3-d Anderson model featuring very weak, uncorrelated disorder, i.e., such that almost all states are non-localized, cf. Sec.~\ref{sec-elimp}. To this end we first determine a pertinent scattering rate matrix. Especially, we choose a specific coarse graining that yields a rate matrix which strictly obeys a certain type of relaxation time approximation (cf. \cite{kittel2004}) and therefore allows for a simple computation of a diffusion coefficient.
The underlying Hamiltonian has the form of Eq.(\ref{Hamiltoneldef}).
The corresponding dispersion relation is given by
\begin{equation}
\varepsilon_{\bf k}=-2J(\cos{k_x}+\cos{k_y}+\cos{k_z})\ ,
\end{equation}
where $J$ is an energy determining the bandwidth (bandwidth=$12J$). Thus the underlying ideal quantum gas corresponds to electrons in a simple cubic lattice.
We focus here on random uncorrelated on-site disorder, i.e., the $W({\bf q})$'s are assumed to be given by independent random numbers, generated according to some distribution (box distribution, gaussian, etc.). As explained below, for this setting only the mean square of the $W({\bf q})$'s eventually enters the rate matrix, thus for our purposes we do not even need to explicitly name the concrete type of distribution here.

\begin{figure}[htb]
\centering
\includegraphics[width=7cm]{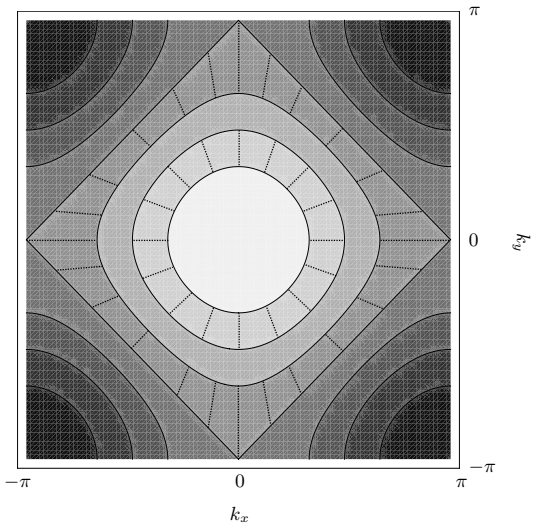}
\caption{Sketch of our specific coarse graining in momentum space for a 2-d Anderson model. Contour lines of equal energy define energy shells of equal energy width $\Delta E$. Within two exemplary shells the full detailed graining is displayed: the shell is orthogonally partitioned into $g_E$ domains of equal area. This amounts to an equal number of states per domain.} \label{fig-disp}
\end{figure}

At this point, we specify the graining in detail. The momentum space is firstly divided into domains corresponding to energy shells of equal energy width $\Delta E$ (according to the dispersion relation). The alignment of the graining with surfaces of constant energy insures that only rates between grains from the same energy shell are non-vanishing, cf. paragraph below (\ref{rateimp}).
Subsequently we further divide the energy shells (labelled by $E$) along the energy gradient into $g_E$ grains, such that each of these grains contains equally many states $N$ which implies equal volumes in momentum space. Fig.~\ref{fig-disp} shows an example of this type of coarse graining for the $2$-dimensional case. We assume that the grains are small enough that the dispersion relation may be linearized on each grain.
Further, we define
\begin{equation}
\overline{W_{\eta\kappa}^{2}}=\sum_{\substack{{\bf i}\in\kappa,\\{\bf k}\in\eta}}\vert W({\bf k}-{\bf i})\vert ^{2}\cdot\frac{1}{N_{\kappa}N_{\eta}}\ .
\end{equation} 
Due to the disorder being uncorrelated we get   
$\overline{W_{\eta\kappa}^{2}}=W^{2}=\text{const.}$, i.e., $ \overline{W_{\eta\kappa}^{2}}$ is independent of the grains. Exploiting this (and as routinely done the properties of the $sinc$-function) we can approximately perform the integration in (\ref{rateimp}), thus finding for times larger than the correlation time (long time limit) the non-diagonal elements of the rate matrix ($\eta\neq\kappa$)

\begin{equation}
R_{\eta\kappa}=\frac{2\pi W^{2}}{\hbar}\frac{N}{\Delta E}\delta_{E(\eta),E(\kappa)}\ ,
\end{equation}
where $\delta_{E(\eta),E(\kappa)}$ is $1$ if $\eta$ and $\kappa$ belong to the same energy shell and $0$ otherwise. So, as already mentioned only grains from the same energy shell are coupled by transition rates. Furthermore, due to the specific coarse graining, these rates are equal for all transitions within one 
energy shell, i.e., there is no specific dependence on $\eta$ and $\kappa$.

The total number of states within some energy shell $E$, which we label by $N_E$, is $N g_E=N_E$.
As explained above, the diagonal terms of the rate matrix may be obtained from the master equation property (Eq.(\ref{diag})).
By means of the abbreviation
\begin{equation}
R^{E}:=\frac{2\pi W^{2}}{\hbar}\frac{N_E}{\Delta E}\ ,
\end{equation}
the rates may be eventually written as
\begin{equation}
R_{\eta\kappa}^{E}=\frac{R^{E}}{g_E}-\delta_{\eta,\kappa}R^{E}\ .
\end{equation}

To repeat, the non-diagonal elements of this rate matrix are all equal and thus, this specific graining yields symmetric detailed balance. As well-known the equilibrium state, i.e., equilibrium set of "grain occupation probabilities" (mathematically: the eigenvector belonging to the eigenvalue $0$) for this type of master equation is the uniform
distribution within one energy shell. So the ``normalized equilibrium vector of probabilities'' corresponding to some energy shell $E$ reads:
\begin{equation}
\vec{P}_0^E: P_{0,\kappa}^E=\frac{1}{g_E},
\end{equation}
for $\kappa$ belonging to the energy shell $E$. Its elements corresponding to all other shells are zero, of course.
Furthermore, for this special type of matrix, all other vectors of occupation probabilities $\vec{X}$, which are orthogonal to equilibrium vectors 
($\vec{X}\vec{P}_0^E=0$) but fall entirely into the respective energy shells $E$, span the (highly degenerate) eigenspaces of the rate matrix with eigenvalues $-R^{E}$. Thus all deviations from equilibrium at energy $E$ relax exponentially with rates $R^{E}$. This is an accurate implementation of the scenario which is assumed to hold approximately if an ``energy dependent relaxation time approximation'' is performed \cite{kittel2004}. 

\begin{figure}[htb]
\vspace{0.4cm}
\centering
\includegraphics[width=6cm]{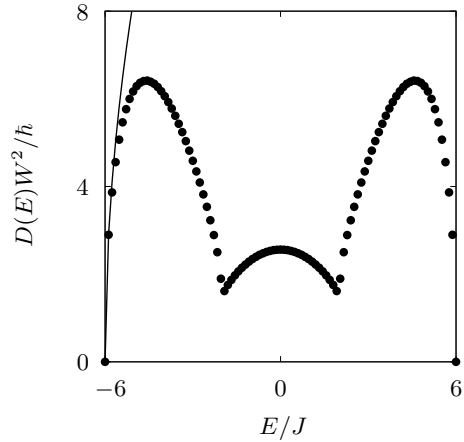}
\caption{Energy dependent diffusion coefficient for a 3-d Anderson model with weak disorder according to formula (\ref{eq-diff}) (dotted curve). The structure is much richer than the one obtained from a free-electron approximation (solid line).}
\label{fig-diff}
\end{figure}

Having found this rate matrix as well as its eigenvectors and eigenvalues we are all set for the calculation of the corresponding diffusion coefficient. There are several approaches to the derivation of a diffusion coefficient from a linear(ized) Boltzmann equation. Here we follow \cite{kadiroglu2007} and references therein by using the formula
\begin{equation}
D_E=-v_{\eta}R^{-1}_{\eta\kappa}v_{\kappa}P^{E} _{0,\kappa} \ ,
\label{difco}
\end{equation}
where $D_E$ is the diffusion coefficient at energy $E$, $v_{\eta},v_{\kappa}$ are the $x$-components (if diffusion in $x$-direction is considered) of the velocities corresponding to the respective grains and $R^{-1}_{\eta\kappa}$ is the (pseudo-)inverted rate matrix (neglecting the null-space). For the velocity components $v_{\kappa}$ we routinely plug in the slope of the dispersion relation in, say, $x$-direction at the respective grains $\kappa$ (group velocity). These slopes are approximately constant within one grain, given one employs the above mentioned graining. Due to the symmetry of the dispersion relation the vector $v_{\kappa}P^{E}_{0,\kappa}$ does not have any overlap with the null-space of the rate matrix, i.e., $\sum_{\kappa}P^{E}_{0,\kappa}v_{\kappa}P^{E}_{0,\kappa}=0$. Hence $R^{-1}_{\eta\kappa}$ from (\ref{difco}) may simply be replaced by $(-1/R^{E})\delta_{\eta\kappa}$, i.e., the inverse of the eigenvalues of the eigenspace that are complementary to the null-space on the corresponding shells.
Hence, we may evaluate formula (\ref{difco}) and eventually find for the diffusion coefficient at energy $E$
\begin{equation}
D_E=\frac{1}{R^{E}}\sum_{\kappa\in E}v_{\kappa}^{2}\frac{1}{g_E}=\frac{1}{R^{E}}\overline{v^{2}}\ .
\label{eq-diff}
\end{equation}
Thus the (shell-specific) diffusion coefficient is completely determined by the total decay rates $R^{E}$ on the shells and the averaged squared velocity in $x$-direction $\overline{v^{2}}$. This expression (\ref{eq-diff}) may easily be evaluated using any standard computer. The result is displayed in Fig.\ref{fig-diff}. 

Obviously the spectrum of diffusion coefficients features a non-trivial structure. Note that the maxima of the diffusion coefficients are not located where the highest $x$-components of the velocities appear, namely in the middle of the spectrum, but more at the edges. In this model the scattering potential features full spherical symmetry, the only anisotropy stems from the dispersion relation. Nevertheless an approximation based on full spherical symmetry yields unacceptable diffusion coefficients for almost the entire spectrum:  A "free-electron approximation", i.e., replacing the true dispersion relation by a pertinent parabolic one (employing some effective mass), makes the calculation even easier (can and has been done analytically \cite{kramer1993}), but describes the diffusion coefficients based on the true dispersion relation only at the outer edges of the peaks more or less correctly, cf. solid line in Fig.\ref{fig-diff}. This insufficiency of spherical approximations is wellknown in the literature and various elaborate ways to cope with it have been developed 
\cite{schoepke1978,papanikolaou1994,dekker1998}. 
With this Anderson model example we intend to demonstrate that the concepts developed in this paper add another straightforward approach to anisotropy.


\section{Summary and Outlook} \label{sec-summary}

We computed linear(ized) master equations describing the dynamics of coarse grained occupation numbers for electrons in periodic lattices experiencing i) elastic scattering, ii) electron-phonon scattering and iii) electron-electron scattering. Our approach 
is based on the time-convolutionless projection operator method. 
The resulting rate matrices are of finite dimension and involve only finite rates. Their derivation does not require the regularization of any singular rates. The concrete form of those rate matrices, including their dimensions, depends crucially on the way the coarse graining is performed (and on the underlying model, of course). They may in general efficiently be numerically evaluated using moderate computing power. A suitable graining may cast the rate matrix in a form that facilitates further considerations which are based on the inversion of the rate matrix such as, e.g., computation of diffusion coefficients. Those inversions may be computed also with moderate computing power, regardless of any symmetry of the underlying model. As an instructive example we presented the calculation of the diffusion coefficient for a 3-d Anderson model with weak, uncorrelated disorder, finding severe deviations from results based on spherical approximations. 
This technique to find convenient representations of collision terms may also serve as a basis for computations of diffusion coefficients in other systems. One may think of electronic diffusion in bulk metals. Pertinent (anisotropic) dispersion relations and scattering potentials could be provided by well-developed numerical methods, such as DFT, etc. and easily incorporated into the approach at hand.  But also electronic diffusion in low dimensional systems such as nanowires embedded in bulk insulators may be addressed as no spherical symmetry is required \cite{bartsch2010}. Furthermore, diffusion coefficients in low dimensional systems controlled by particle-particle interactions may be addressed (cf. \cite{sirker2009}). If eventually conductivity rather than the diffusion coefficient is the quantity of interest, diffusion coefficients may be converted to conductivities using a generalized Einstein relation (cf. \cite{steinigeweg2009}). Thus the results presented here may be used as a basis for a range of  concrete transport-theoretical investigations. The feasibility of such an approach is supported by the fact that a similar approach has been demonstrated to yield reasonable results for electronic lifetimes in aluminum \cite{kadiroglu2009}.

%
%

\begin{acknowledgments}

We sincerely thank M. Kadiro\=glu for his contributions to fruitful
discussions. Financial support by the ``Deutsche
Forschungsgemeinschaft'' is gratefully acknowledged.

\end{acknowledgments}

%
%

\begin{appendix}

\section{Traditional Approaches to Quantum Boltzmann Equations} \label{sec-app}

Just to give some background information we intend to very briefly (and incompletely) review here the various suggestions for mappings of quantum dynamics of non-ideal quantum gases onto Boltzmann equations that may be found in the literature. To those ends we classify them as three main groups of approaches: i) ''Fermi's Golden Rule approaches'', ii) ''Green's-functions approaches'' and iii) ''factorization approaches''. Of course there are many interconnections between them but for clarity we keep to this scheme here. 

i): In this rather phenomenological approach the absolute square of the interaction matrix element that ''connects'' two occupation number eigenstates is boldly taken as the weight of a singular, classical transition rate between those two states, which already implies a kind of random phase approximation \cite{peierls2001}. Furthermore, it is assumed that the probabilities of the occupation number eigenstates are and remain such that occupation numbers factorize. This leads to an autonomous equation of motion in terms of mean occupation numbers (see also \cite{peierls2001,lepri2003,jaeckle1978}).

ii): Here the issue is approached by setting up the hierarchy of equations of motion for the Green's functions. The higher order Green's functions are then re-expressed by two-point-Green's functions using diagrammatic techniques and corresponding approximations. This results in an autonomous equation of motion in terms of one-mode-two-time-Green's functions. The latter are directly interpretable as mean occupation numbers for the case of full equilibrium only. However, to leading order this approach yields an outcome comparable to the result from i). For details see, e.g., \cite{kadanoff1962,rammer1968,zubarev1996}.

iii): The starting point are the Heisenberg equations of motion for the occupation numbers. Iterating those w.r.t. time and assuming full factorizability and diagonality of the occupation number operators at all times yields an autonomous set of equations for the mean occupation numbers. Again, this approach yields an outcome comparable to the result from i). For details see, e.g., \cite{erdos2009,haug2006,banyai2006}.

All of the above considerations are based on discrete momentum lattices. Within the frame of such a description the transition rates become singular if the infinite time limit is considered, which is what is usually done. These singularities have to be regularized ``by hand'' before the description may be transferred to a continuous momentum space. 

(Of course there are much more investigations of Boltzmann equations for quantum systems, some explore yet different approaches to non-ideal quantum gases such as, e.g.,\cite{aoki2006,koide2000,benedetto2004}, some treat other systems, e.g., \cite{hornberger2006,vacchini2005,kadiroglu2007}, etc..)

\end{appendix}

%
%

\end{document}